\documentclass[12pt, centerh1]{article}
  \usepackage{metalogo} 
  \usepackage{algpseudocodex} 

  \usepackage{color}
  \usepackage{array}
  \usepackage{soul}
  \usepackage{enumitem} 
  \usepackage{bm}
  \usepackage{amsmath}
  \usepackage{natbib}
  \usepackage{hyperref}
  \usepackage{amssymb}
  \usepackage{amsthm}
   \usepackage{subcaption}
\usepackage{float}
\usepackage{mathrsfs}
 \usepackage{graphicx}
  \graphicspath{%
    {converted_graphics/}
    {Figures-2016-06-13/}
}

\definecolor{jade}{rgb}{0.0, 0.66, 0.42} 

\newcommand{\comR}[1]{\textcolor{black}{#1}}

  \title{Weighted Parameter Estimators of the Generalized Extreme Value Distribution in the Presence of Missing Observations}

  \author{James H. McVittie\textsuperscript{1} and Orla A. Murphy\textsuperscript{2}*}
  \date{\small \textsuperscript{1}Department of Mathematics and Statistics, University of Regina, Saskatchewan, Canada\\
  \textsuperscript{2}Department of Mathematics and Statistics, Dalhousie University, Nova Scotia, Canada\\
  *Corresponding author. E-mail: orla.murphy@dal.ca}

\begin{document}
\maketitle
\begin{abstract}
Missing data occur in a variety of applications of extreme value analysis. In the block maxima approach to an extreme value analysis, missingness is often handled by either ignoring missing observations or dropping a block of observations from the analysis. However, in some cases, missingness may occur due to equipment failure during an extreme event, which can lead to bias in estimation. In this work, we propose weighted maximum likelihood and weighted moment-based estimators for the generalized extreme value distribution parameters to account for the presence of missing observations. We validate the procedures through an extensive simulation study and apply the estimation methods to data from multiple tidal gauges on the Eastern coast of Canada. \\ \vspace{0.5cm}

\end{abstract}
\newpage

\section{Introduction}

The analysis of extreme events, which are both rare and large in magnitude, is a challenging modelling problem. The presence of missing observations further complicates the extreme value analysis as the extreme events may not be observed or known to be missing. Missing observations in an extreme value analysis occur frequently in a variety of applied settings. For example, in an environmental application where the quantity of interest is the height of coastal wave surges, extreme events can lead to equipment failure in the buoys or tidal gauges resulting in missing data records \cite{ayiah}. Consequently, analyses ignoring missing observations may yield biased results and increase the overall variability in the computed estimates. To mitigate such issues, it is critical to account for missing observations in the proposed analyses. 
\

Various techniques have been developed to analyze extreme values in the presence of missing observations. In \cite{beirl, xu, zou}, authors examined the Hill estimator under various missing data and distributional assumptions. Specifically, these works considered the missing data scenario where the largest order statistics in the observed data set were missing. In contrast, \cite{ayiah} and \cite{turki2020} both considered different types of multiple imputation to fill in the missing observations in their extreme value analysis. This multiple imputation approach implicitly assumes the probability of an observation being missing is completely independent of the observed data or depends only on the observed data In \cite{ryden}, the author examined the missing data problem from an applied perspective by studying return values of wave height whereas in \cite{glava}, the authors examined the asymptotic properties of moving averages from the Gumbel distribution in the presence of missing observations. Alternatively, the analysis of \cite{beck} simply removed the blocks of observations with high levels of missingness according to some preset threshold and treated all other maxima as though they were derived from fully observed blocks of data. While this approach seems to be used more often in practice, there is no general consensus on an optimal technique for handling missing records.    
\ 

In this work, we borrow approaches from robust estimation to propose weighted estimation methods for the block maxima approach of modeling extremes. Specifically, we consider weighted likelihood and moment estimators to account for missing observations within blocks. In Section \ref{sec:BMest}, we review two unweighted estimation procedures (maximum likelihood estimation, probability-weighted moment estimators) commonly used for the block maxima approach when estimating the generalized extreme value distribution parameters. We formally define the ways in which observations may be missing in Section \ref{sec:WE} and discuss their relevance in applications related to extreme wave surges. To account for missingness within blocks, we also introduce weights into the maximum likelihood and probability-weighted moment estimators where the weights are either determined unconditionally/conditionally on the observed data in Section \ref{sec:WE}. In Section \ref{sec:SS}, we conduct a simulation study to compare the performance of the various estimators in the presence of missing values. We apply our proposed weighted estimators to three tidal gauge stations in Eastern Canada to estimate the 20-, 50- and 100-year return levels in Section \ref{sec:data} and we conclude with a short discussion of future research directions in Section \ref{sec:disc}.

\section{Block Maxima Approach to Extremes}\label{sec:BMest}

The block maxima approach to extreme value analysis divides a series of data into blocks of equal size. In this approach, only the maximum within each block is considered as an extreme value and used to fit an extreme value model. There is often a natural choice of block size, e.g., monthly or yearly blocks, where the relevant extreme value model, defined below, is the asymptotic model as the block size approaches infinity. 

Let $X_1, ..., X_N$ denote a random sample drawn from CDF $F(\cdot)$ where $N$ is partitioned into $k$ blocks each of size $n$, such that $kn = N$. For each $j \in \{1, 2, ..., k\}$, let $M_{j}=\max(X_{j, 1},\dots,X_{j,n})$ denote the $j$th block maximum and denote a general maximum of $n$ observations by $M_n$, i.e., $M_n$ equals the $j$th block maxima $M_j$ in distribution. Suppose there exists sequences of constants $\{a_n>0\}$ and $\{b_n\}$ such that, for all $z\in\mathbb{R}$,
\begin{equation}\label{eq:GEVlim}
\lim_{n\rightarrow\infty} P\left(\frac{M_{n}-b_n}{a_n}\leq z \right) = G(z)
\end{equation}  
for some non-degenerate distribution $G$. Then $G$ is the generalized extreme value (GEV) distribution:
\begin{equation}
G(z; \mu, \xi, \sigma)=\begin{cases} 
\exp \left[ - \left\{ 1 + \xi \left( \frac{z - \mu}{\sigma} \right) \right\}^{-1/\xi} \right]&\text{ if } 1 + \xi(z-\mu)/\sigma > 0,\xi\neq0, \\
\exp \left\{ - \exp\left(  \frac{z - \mu}{\sigma} \right)\right\}&\text{ if }z\in \mathbb{R}, \xi=0, \\
\end{cases}
\label{gev}
\end{equation}  
where $-\infty < \mu < \infty$, $\sigma > 0$, and $-\infty < \xi < \infty$ \cite{fishe,gnedenko1943,mises1936}. Note that this result also holds for certain stationary series, specifically those satisfying the condition that the form of the temporal dependence results in extreme events that are approximately independent if they are sufficiently far apart. See \cite{leadbetter1974extreme} for relevant theoretical results and \cite{beirlant2004} for an in-depth explanation of the extension of this result to stationary series and estimation.

In practice, we approximate the distribution of $M_n$ by $G$ for large enough $n$. The parameters of the GEV distribution are typically estimated by one of two methods: (i) maximum likelihood (ML) estimation or (ii) probability weighted moment estimators (PWM). For the sake of simplicity in the estimation descriptions below, we suppose the maxima $M_{1}$, ..., $M_{k}$ are independently and identically distributed according to a GEV distribution. 

\subsection{Maximum Likelihood Estimation}

The ML estimator is a set of parameter estimators which maximize the likelihood function, or equivalently the log likelihood function, using the realizations $m_{1}, ..., m_{k}$:  
\begin{equation}
\hat\Theta = {\arg\max}_\Theta \ell(\Theta;m_{1}, ..., m_{k}) = {\arg\max}_\Theta \sum_{j=1}^{k} \log \left\{ g(m_{j}; \Theta) \right\},
\label{like}
\end{equation}
where $g$ is the GEV density function and $\Theta = \{\mu, \xi, \sigma\}$. In practice, the convergence result in \eqref{eq:GEVlim} ensures the maxima are approximately distributed according to the generalized extreme value distribution for large enough block size, $n$. As described in \cite{smith}, if the shape parameter $\xi > -1$, then the ML estimators for $\sigma$, $\mu$ and $\xi$ are consistent:
$$(\hat{\sigma}, \hat{\mu}, \hat{\xi}) \xrightarrow{\mathbb{P}} (\sigma, \mu, \xi),$$
as $k \rightarrow \infty$ and if $\xi > -0.5$, the ML estimators are also asymptotically Normally distributed:
$$\sqrt{m}\{(\hat{\sigma}, \hat{\mu}, \hat{\xi})\} \xrightarrow{\mathcal{D}} N(0, V),$$
where $V$ is the inverse Fisher information matrix \cite{smith}. Recently, a refinement and completion of the proof of \cite{smith} was given in \cite{zhang}.

\subsection{Probability-Weighted Moment Estimation}

An alternative to the maximum likelihood estimation approach is the PWM method \cite{green, hoski}. For $p,r,s\in\mathbb{R}$, define the probability weighted moments by:
\begin{equation}
M_{p,r,s}=E\{X^p[F(X)]^r[1-F(X)]^s \}.
\label{moment}
\end{equation}
As above, let $M_{1},\dots,M_{k}$ denote a random sample from the GEV distribution where $\xi\in(-\infty,1)\setminus\{0\}$. Then, for $p=1$, $r=0,1,2,\dots$, and $s=0$, it can be shown that:
\begin{equation}
M_{1,r,0}=\frac{1}{r+1}\left\{\mu-\frac{\sigma}{\xi}[1-(r+1)^\xi \Gamma(1-\xi)]\right\}.
\label{pwm}
\end{equation}
From the expectation given in Equation \eqref{moment}, an unbiased estimator for ${M}_{1,r,0}$ is given by:
\begin{equation}
\hat{M}_{1,r,0} = \frac{1}{k} \sum_{j=1}^{k} \left( \prod_{l=1}^{r} \frac{j-l}{k-l} \right) M_{j}.
\label{mom}
\end{equation}
The moment estimators yield a system of equations which can be solved for the unknown parameters of the GEV distribution:
\begin{equation}
\begin{cases}
\hat{\sigma} = \frac{\hat{\xi} (2 \hat{M}_{1,1,0} - \hat{M}_{1,0,0})}{\Gamma(1 - \hat{\xi})(2^{\hat{\xi}} - 1)} \\
\hat{\mu} = \hat{M}_{1,0,0} + \frac{\hat{\sigma}}{\hat{\xi}} (1 - \Gamma(1 - \hat{\xi})) \\
\frac{3\hat{M}_{1,2,0} - M_{1,0,0}}{2\hat{M}_{1,1,0} - \hat{M}_{1,0,0}} = \frac{3^{\hat{\xi}} - 1}{2^{\hat{\xi}} - 1}, \text{ where }\hat{\xi} \text{ is solved numerically.}
\end{cases}
\label{momests}
\end{equation}
In addition to deriving the PWM estimator, it can be shown that the PWM estimators are consistent and asymptotically Normally distributed for $\xi<0.5$ \cite{hoski}. Furthermore, the PWM type estimators can outperform likelihood-based estimators for the GEV distribution parameters in cases of small sample sizes (i.e. small values of $k$) \cite{hoski}. This quality will be highlighted in our estimates of return levels of wave surges using data collected from tidal gauges located on the Eastern coast of Canada. 

\section{Weighted Estimation Approaches}\label{sec:WE}

\begin{figure}
\centering
\includegraphics[scale=0.45]{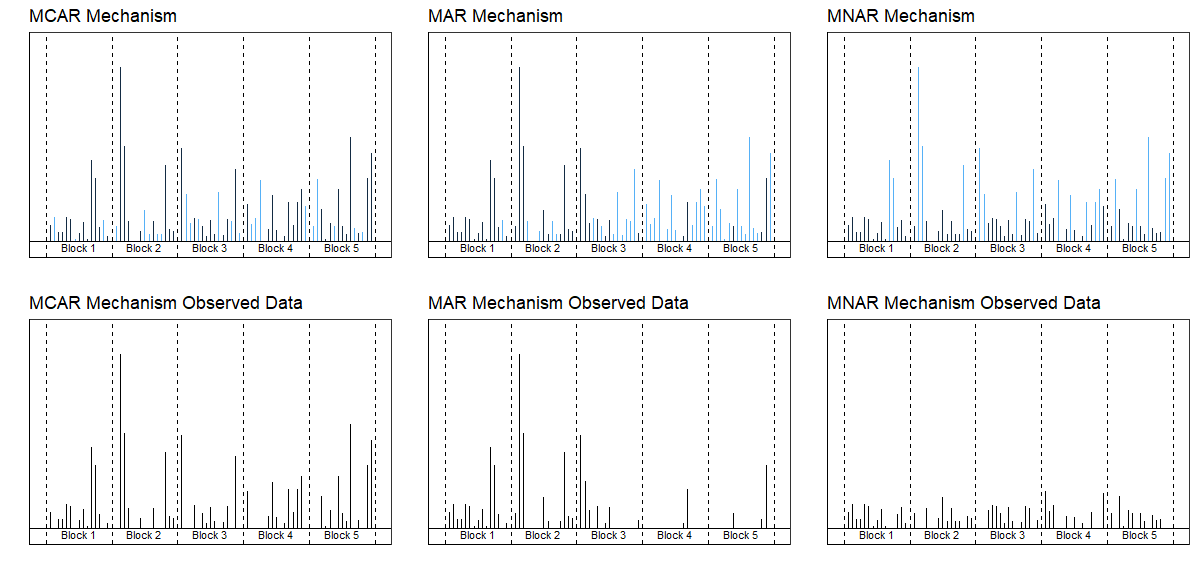}
\caption{A graphical representation of the effects of different missingness mechanisms on block maxima data. The blue lines in the upper three panels represent the missing observations. The lower three panels only include the observed data in each of the blocks.}
\label{missdatplot}
\end{figure}

In the presence of missing data, the ML and PWM estimators can yield biased estimates as the observed block maxima may not correspond to the true block maxima across multiple blocks. To adjust for missing observations, we propose weighted maximum likelihood and weighted moment-based estimators to estimate the GEV model parameters. 
\

\subsection{Missingness in the block maxima framework}

First, we slightly modify the previous notation to account for missing observations. Suppose of the series of independent and identically distributed $N$ random variables, $N_{obs}$ are observed and $N_{miss}$ are missing (i.e. $N_{obs} + N_{miss} = N$). Let $\mathscr{S}$ and $\mathscr{S}'$ denote the disjoint sets of indices of sizes $N_{obs}$ and $N_{miss}$ for the observed and missing random variables, respectively, such that $\mathscr{S}_N=\mathscr{S} \cup \mathscr{S}'=\{1,\dots,N\}$. We further let $\{X_{i}:i\in\mathscr{S}\}$ and $\{X_{i}':i\in\mathscr{S}'\}$ denote the corresponding sets of observed and missing random variables all with CDF $F(\cdot)$. Again, we suppose $N$ is partitioned into $k$ blocks each of size $n$, $kn = N$. In the $j$th block, denote the maximum based on the $n_j$ observed and $n_j'$ missing observations as $M_{n_j}$ and $M_{n_j'}$, respectively, and the true block maximum as $M_j$, where $n_j+n_j'=n$. 
\

The underlying mechanism for missingness can vary across datasets, but these mechanisms generally fall into three categories: (i) \emph{missing completely at random} (MCAR), (ii) \emph{missing at random} (MAR) and (iii) \emph{missing not at random} (MNAR). In the case of MCAR, the probability of missingness is independent of both the observed and unobserved data. In the case of MAR, the probability of missingness is dependent only on the observed data. If the first two categories do not hold and the probability of missingness is dependent on unobserved data, then the missingness mechanism falls into the category of MNAR. See \cite{littl} for further details on the different missingness mechanisms.
\

Using the missingness notation within the block maxima framework, the three categories of missingness can be described probabilistically as follows. Suppose $I$ is an indicator of missingness for an arbitrary random variable $X$, where $I=1$ if $X$ is observed and 0 otherwise.
\begin{enumerate}
\item[MCAR:] The probability of missingness is fixed across all observations in the series, i.e., $P(I=0\mid\{X_{i}:i\in\mathscr{S}_N\})=P(I=0)$. In this case, each block may or may not have a portion of missing observations.
\item[MAR:] The probability of missingness depends only on the observed random variables, i.e.,\\ $P(I=0\mid\{X_{i}:i\in\mathscr{S}_N\})=P(I=0\mid\{X_{i}:i\in\mathscr{S}\})$.
\item[MNAR:] The probability of missingness depends on unobserved random variables, i.e., $P(I=0\mid\{X_{i}:i\in\mathscr{S}_N\})$ cannot be simplified to remove $\{X_{i}':i\in\mathscr{S}'\}$ from the conditioning.
\end{enumerate}
\

In practice, it is more likely to encounter missingness mechanisms that fall into the MAR and MNAR categories, rather than MCAR. In the context of wave surges, missingness being completely independent of wave intensity or any other information from the observed/unobserved data would indicate MCAR. In contrast, increased failure of the measuring equipment over time would be an example of MAR, and failure of equipment to measure high intensity surges would be an example of MNAR. In Figure \ref{missdatplot}, we illustrate how different missingness mechanisms affect the observed block maxima data. Specifically, we generated 75 draws from an Exponential distribution with rate parameter $0.2$ and applied either an MCAR, MAR or MNAR missingness mechanism in the left, middle and right panels, respectively. In the left most panels, we assumed an MCAR missingness mechanism where series level data were missing according to independent Bernoulli draws with probability equal to $0.3$. Under this missingness mechanism, a particular block's maximum will be missing with probability $0.3$. In the lower left panel of Figure \ref{missdatplot}, blocks 1 to 3 and 5 still contain their true maxima whereas block 4's observed maxima now corresponds to the second largest maxima in the block. Similarly, in the middle panels, we imposed a MAR mechanism where the series level are missing with increasing probability according to their time indices (i.e. an observation's probability of being missing increases according to a Normal cumulative distribution function). Thus, the earlier blocks have fewer missing observations whereas later blocks have mostly missing observations. The corresponding effect on an extreme value analysis is that the observed maxima in earlier blocks are more likely to correspond to the true block maxima whereas later blocks' observed maxima will correspond to the lower order statistics. In contrast, in the right most panels of Figure \ref{missdatplot}, we consider the most extreme case of a MNAR mechanism where the highest 25\% of the order statistics of the entire series are removed. This scenario is the most destructive to an extreme value analysis as the upper order statistics on which an analysis is based are totally unobserved. In this example, all five blocks' observed maxima correspond to much lower order statistics and will therefore yield biased inferences. We note that in any applied analysis, certain assumptions are required on the type of missingness mechanism. Thus, the described missingness mechanisms, while not necessarily applicable to all real world data scenarios, will be considered in an extensive simulation study in Section 4 as possible cases on a spectrum of how negatively missing series level data can impact block maxima inferences.
\

To handle missing data under the block maxima approach for the ML or PWM estimators, we introduce weights $\hat{w}_j$ for each block $j = 1, ..., k$ where $0 \leq \hat{w}_j \leq 1$ such that higher weights are assigned to block maxima where the true block maxima was more likely to have been observed. Specifically, we propose two weighting schemes:
\begin{equation}
\hat{w}_j^{(1)} = \frac{n_j}{n_j + n_j'} \quad \text{ and } \hat{w}_j ^{(2)}= \left\{ \frac{1}{N_{obs}} \sum_{i=1}^{N_{obs}} I(X_i \leq m_{n_j}) \right\}^{n_j'} = \hat{F}_{N_{obs}}(m_{n_j})^{n_j'}.
\label{wgts}
\end{equation}
We denote $\hat{w}_j^{(1)}$ as the unconditional estimate of not missing the true block maximum within block $j$, i.e., $\hat{w}_j^{(1)}=\hat{P}\{M_{j}=M_{n_j}\}$, as it depends only on the number of observations (observed/missing) within block $j$ and not on block $j$'s data. In contrast, we denote $\hat{w}_j^{(2)}$ as the conditional estimate of not missing the true block maximum within block $j$, i.e. $\hat{w}_j^{(2)}=\hat{P}(M_{j}=M_{n_j}\mid M_{n_j}=m_{n_j}, \{X_i\}_{i\in\mathscr{S}})$, as it depends on all non-missing observations across all blocks. Throughout the remainder of the article, we will assume that any analyses are conducted conditionally on the number of missing and non-missing observations within each of the blocks. Additional commentary regarding unconditional analyses is included in Section 6.
\

We note that although these weighting schemes are inspired by MCAR and MAR missingness mechanisms, respectively, their use is not restricted to such cases as these missingness mechanisms are not required as assumptions for deriving the properties of the corresponding estimators outlined in Section \ref{subsec:prop}. As such, the definitions of the unconditional/conditional weights were proposed in order to account for the three missingness categories. We assess the performance of the proposed weighted estimation procedures when applied to simulated data under three missingness mechanisms in \ref{sec:SS}.


\subsection{Weighted Maximum Likelihood Estimation}

Weighted ML estimation is a technique of downweighting the effect of outlying observations in an extreme value analysis (see \cite{carreau2011, dupuis2002, wang2010} for further details) as well as accounting for missingness through inverse probability weighting \cite{seaman2013}. The weighted log-likelihood of the GEV model is defined for an appropriate choice of weights $\hat{w}_j$ for $j=1,\dots,k$ as follows:
\begin{equation}
\ell_{\hat{\boldsymbol{w}}}(\Theta;m_{n_1}, ..., m_{n_k}) = \sum_{j=1}^{k} \hat{w}_j \log \left\{ g(m_{n_j}; \Theta) \right\},
\label{wgtlike}
\end{equation}
where $w_j \in[0,1]$. In our setting, the weights would be first estimated either unconditionally/conditionally and then substituted into Equation \eqref{wgtlike} through which estimates of $\Theta$ would be obtained through a standard optimization procedure. For extensions to weighted maximum likelihood estimators in the bivariate extreme model setting, see \cite{dupui}.

\subsection{Weighted Probability-Weighted Moments Estimator}

Similar to the weighted ML estimation procedure, we propose a weighted PWM estimator to account for missingness within blocks by incorporating weights $\hat{w}_j$ within $\hat{M}_{1,r,0}$ in \eqref{momests} in the following way:
\begin{equation}
\hat{M}^{\hat{\boldsymbol{w}}}_{1,r,0} = \frac{1}{\sum_{j=1}^{k} \hat{w}_j} \sum_{j=1}^{k} \hat{w}_j \left( \prod_{l=1}^{r} \frac{j-l}{k-l} \right) M_{n_j}.
\label{wgtmom}
\end{equation}
As in the ML procedure, we first determine the weights from Equation \eqref{wgts} and then substitute these into Equation \eqref{wgtmom} to obtain a weighted estimator for $M_{1,r,0}$. The corresponding GEV distribution parameters are then determined through the same update equations given in Equations \eqref{momests}.

\subsection{Statistical properties of estimators}\label{subsec:prop}

As described in Section 2, when there are no missing observations, the parametric maximum likelihood and method of moment estimators are consistent and asymptotically Normally distributed. In the presence of missing observations, depending on the types of estimated weights, the weighted estimators can exhibit similar properties.
\

For the weighted likelihood estimators, it is not immediately clear whether the weighted ML estimator for the GEV parameters will be consistent and asymptotically Normally distributed. The log-likelihood in Equation \eqref{wgtlike} can be regarded as a relevance weighted log-likelihood function as the observed block maxima are not necessarily drawn from the GEV distribution \cite{hu}. The associated weights correspond to the relevance of the observed block maxima densities relative to the true block maxima density (i.e. the GEV distribution). There are multiple challenges associated in establishing the asymptotic properties of the maximum relevance weighted likelihood estimator. Depending on the assumptions underlying the missingness mechanism, it is not evident how to define the asymptotic distribution of the observed maxima as the number of observations within the block tends to infinity. Specifically, the relative position of the observed block maxima (i.e. the order of the block's largest observed order statistic) will be dependent on the growth rate of the number of observed/unobserved random variables within the block. Furthermore, when the weights are estimated conditionally and evaluated at the observed maxima, $\hat{w}_j^{(2)}=\hat{P}(M_{j}=M_{n_j}\mid M_{n_j}=m_{n_j}, \{X_i\}_{i\in\mathscr{S}})$, the asymptotic arguments of \cite{hu} do not immediately follow as the weights are random. An analogous data-dependent weighting approach was recently considered by \cite{biswa} however the corresponding asymptotic properties have yet to be determined. A mathematical description highlighting some of the challenges stated above and the required regularity conditions stated by \cite{hu} for consistency and asymptotic normality is given in Appendix A.  
\



In the case of the weighted method of moments estimators, some comparisons are possible between the probability-weighted moment estimators when all block maxima are correctly observed and the weighted probability-weighted moment estimators when some observations are missing. As discussed in \cite{beirlant2004}, the probability-weighted moment estimator given in Equation \eqref{mom} is unbiased for $M_{1,r,0}$. When $\hat{M}_{1,r,0}$ is defined in terms of the observed maxima given by $M_{n_j}$, since $M_{n_j} \leq M_j$, then $\mathbb{E}(\hat{M}_{1,r,0}) \leq M_{1,r,0}$. Similarly, for the weighted probability-weighted moment estimators, since $\hat{w}_j^{(1)}/\sum_{j=1}^{k} \hat{w}_j^{(1)} \leq 1$ and $\hat{w}^{(2)}_j/\sum_{j=1}^{k} \hat{w}_j^{(2)} \leq 1$, for all $j = 1, ..., k$, then $\mathbb{E}(\hat{M}^{\hat{\boldsymbol{w}}}_{1,r,0}) \leq M_{1,r,0}$. The asymptotic Normality result of \cite{hoski} requires little modification when applied to $\hat{M}^{\hat{\boldsymbol{w}}}_{1,r,0}$ using the unconditional weights as the proof techniques are based on the work of \cite{chern} which allows for arbitrary weights that are functions of $j$ and $n$. For further details on how the assumptions in the proof of \cite{chern} still hold with the inclusion of unconditional weights, see Appendix A. However, as with the weighted maximum likelihood estimators, further research is required to derive the asymptotic properties of the weighted probability-weighted moment estimators where the weights are dependent on the observed block maxima.   

\section{Simulation Study}\label{sec:SS}

To assess the performance of the estimation procedures and different weight functions, we simulated block maxima in a variety of different scenarios. The data generating process was on the observational scale and considered to be from either the Exponential, Student t or Beta distributions, which belong to the Gumbel,  Fr\'echet and Weibull maximum domains of attraction, respectively. We considered block sizes of 50 or 100 and varied the number of blocks from 25 to 100 by increments of 25. For each number of blocks and block size, we considered different percentages of blocks with missingness and different levels of missingness within blocks. 
\

In this work we consider three missingness mechanisms, one relating to each category of missingness. For the first case, MCAR, the missing observations are selected at random with a fixed probability within each block. For the second case, MAR, the probability of missingness depends only on the time index and becomes more likely as time progressed. For the third case, MNAR, following the work of \cite{beirl} and \cite{einmahl2008}, we removed the higher order statistics within blocks according to the percentage of missing observations. Thus, if a block has missing observations, then its true maximum is necessarily missing. We note that these missingness mechanisms are of varying levels of plausibility for wave surge data but they are used with the intent of evaluating robustness of the proposed methods to different mechanisms and therefore help evaluate the broader performance of the proposed methods. Specifically, we use these types of missingness mechanisms as references on the spectrum of the ways in which the block maxima can be potentially unobserved. As the MCAR missingness mechanism does not necessarily imply loss of block maxima, it can be regarded as the least damaging to the extreme value analysis, whereas the MNAR missingness mechanism, which assumes the extreme order statistics are missing, can be regarded as the most damaging. In practice, a missingness mechanism may not fall into these extreme cases and so we also consider a MAR missingness mechanism which depends on the observed data.   
\

We replicated the simulation procedure over $1000$ iterations and computed the unnormalized Cram\'er--von Mises (CvM) distances, where the CvM distance is defined by:
$$\text{CvM} = \int \left[\hat{G}(x) - G(x)\right]^2 \text{d}G(x),$$
in which $G(x)$ corresponds to the estimated GEV distribution if all observations across blocks were known and $\hat{G}(x)$ corresponds to the estimated GEV distribution using one of the proposed estimation methods from Section 3. The CvM distance is approximated by the summation:
$$\frac{1}{N^\ast} \sum_{i=1}^{N^\ast} \left[\hat{G}(x_i) - G(x_i)\right]^2,$$
where $x_i$ corresponds to the $N^\ast$ incremental percentiles of $G(\cdot)$ from $2\%$ to $98\%$, which yields similar results to a Monte Carlo-based estimate. 

We report the t-distribution simulation results in Tables \ref{simtablet-MNAR}-\ref{simtablet-MAR}, and the Exponential and Beta distribution simulation results in Tables 1-6 in the Supplementary Materials. From Tables \ref{simtablet-MNAR}-\ref{simtablet-MAR}, the conditional-weighted ML estimation outperforms the unweighted and unconditionally weighted ML methods in all scenarios, i.e., the conditional ML estimator yielded the lowest CvM distances irrespective of whether the data were MNAR, MCAR or MAR. However, in cases of high levels of missingness, e.g., when the percentage blocks with missing observations is 80\%, the conditional ML estimator tended to perform poorly. In the case of the t distribution, we found no changes in the reported simulation results when the degrees of freedom of the t distribution increased from $5$ to $15$ to $40$ (i.e. as the distribution tails become lighter). The superiority of the conditional ML estimator compared to the observed and unconditional ML estimators is most likely attributable to the improved accuracy of the weight in modelling the missingness probability. Specifically, since the conditional weights are based on the series level data, the empirical CDF will yield a consistent and unbiased estimator for whether the true block maxima are missing. As the unconditional weights are based solely on the relative sizes of the observed to unobserved data within the blocks, they do not utilize any information from the series level data nor the observed maxima. The results from the PWM-based estimation are less consistent. Although the weighted PWM-based estimation generally outperforms the unweighted PWM-based estimation in the MNAR setting, there is no clear improvement in performance between the unconditional and conditional weights. In the MAR and MCAR settings, this pattern is reversed with the unweighted PWM-based estimator outperforming the weighted PWM-based estimators. Furthermore, the conditional weighted likelihood method outperforms the PWM-based estimation methods in most cases irrespective of the data generating distribution or the missingness mechanism for the  simulated data. We conjecture that the weighted moment estimators' poor behaviour is due to the reweighting of the observed maxima rather than the likelihood contributions as in the ML estimation procedure. By reweighting the observed maxima, the weighted PWM-based estimators are attempting to correct for the cases where the observed block maxima do/do not correspond to the true block maxima values. Prediction of new extreme order statistics based on the observed order statistics is highly variable and thus can lead to innaccurate predictions. Based on the stability of the ML estimator simulation results given above relative to the PWM estimator results, we would advise using the conditionally weighted ML estimator to obtain estimates of the GEV parameters when series level observations are missing.

\begin{table}[!htbp] \centering \footnotesize
  \caption{Average empirical Cram\'er--von Mises distances (non-normalized) between the CDFs of the estimated complete data GEV and the proposed modelling approach computed over $1000$ replications (``mle-obs'' - observed likelihood, ``mle-uncond'' - unconditional weighted likelihood, ``mle-cond'' - conditional weighted likelihood'', ``mom-obs'' - observed moments, ``mom-uncond'' - unconditional weighted moments, ``mom-cond'' - conditional weighted moments) for MNAR data generated from an t distribution (df=5) with blocks containing $100$ observations. Simulation parameters: ``sims'' - number of blocks, ``pbm'' - proportion of blocks with missing observations, ``pm'' - expected proportion of missing observations within blocks.} 
    \label{simtablet-MNAR} 
\begin{tabular}{@{\extracolsep{5pt}} ccc|ccc|ccc} 
\\[-1.8ex]\hline 
\hline \\[-1.8ex] 
sims & pbm & pm & mle-obs & mle-uncond & mle-cond & mom-obs & mom-uncond & mom-cond \\
\hline \\[-1.8ex] 
$25$ & $0.200$ & $0.050$ & $15.594$ & $14.363$ & $\textbf{9.861}$ & $14.622$ & $13.752$ & $\textbf{11.726}$ \\ 
$50$ & $0.200$ & $0.050$ & $14.252$ & $13.101$ & $\textbf{8.759}$ & $12.847$ & $12.162$ & $\textbf{10.566}$ \\ 
$100$ & $0.200$ & $0.050$ & $13.715$ & $12.587$ & $\textbf{8.237}$ & $11.862$ & $11.234$ & $\textbf{9.817}$ \\ 
$25$ & $0.500$ & $0.050$ & $86.768$ & $82.387$ & $\textbf{67.015}$ & $74.178$ & $69.181$ & $\textbf{48.846}$ \\ 
$50$ & $0.500$ & $0.050$ & $84.492$ & $80.057$ & $\textbf{64.116}$ & $70.795$ & $65.929$ & $\textbf{45.913}$ \\ 
$100$ & $0.500$ & $0.050$ & $83.510$ & $79.062$ & $\textbf{63.152}$ & $69.542$ & $64.619$ & $\textbf{44.406}$ \\ 
$25$ & $0.800$ & $0.050$ & $187.942$ & $184.224$ & $\textbf{172.502}$ & $176.415$ & $169.435$ & $\textbf{130.062}$ \\ 
$50$ & $0.800$ & $0.050$ & $185.762$ & $181.831$ & $\textbf{169.656}$ & $174.120$ & $167.047$ & $\textbf{126.731}$ \\ 
$100$ & $0.800$ & $0.050$ & $183.270$ & $179.218$ & $\textbf{166.789}$ & $172.029$ & $164.916$ & $\textbf{123.859}$ \\ 
$25$ & $0.200$ & $0.200$ & $25.599$ & $18.801$ & $\textbf{2.489}$ & $23.783$ & $\textbf{19.785}$ & $23.175$ \\ 
$50$ & $0.200$ & $0.200$ & $24.311$ & $18.081$ & $\textbf{1.377}$ & $21.166$ & $\textbf{17.869}$ & $22.213$ \\ 
$100$ & $0.200$ & $0.200$ & $24.342$ & $18.170$ & $\textbf{0.917}$ & $20.130$ & $\textbf{17.037}$ & $22.056$ \\ 
$25$ & $0.500$ & $0.200$ & $135.521$ & $112.493$ & $\textbf{14.065}$ & $116.162$ & $96.553$ & $\textbf{50.550}$ \\ 
$50$ & $0.500$ & $0.200$ & $140.356$ & $116.329$ & $\textbf{9.946}$ & $114.339$ & $94.280$ & $\textbf{47.881}$ \\ 
$100$ & $0.500$ & $0.200$ & $140.500$ & $116.457$ & $\textbf{8.502}$ & $111.756$ & $91.509$ & $\textbf{45.570}$ \\ 
$25$ & $0.800$ & $0.200$ & $253.488$ & $239.038$ & $\textbf{148.422}$ & $248.220$ & $233.501$ & $\textbf{158.392}$ \\ 
$50$ & $0.800$ & $0.200$ & $255.664$ & $240.082$ & $\textbf{144.716}$ & $245.631$ & $230.519$ & $\textbf{152.088}$ \\ 
$100$ & $0.800$ & $0.200$ & $257.355$ & $241.139$ & $\textbf{143.399}$ & $245.711$ & $230.209$ & $\textbf{149.443}$ \\ 
$25$ & $0.200$ & $0.350$ & $27.091$ & $16.160$ & $\textbf{1.781}$ & $25.482$ & $\textbf{19.789}$ & $23.881$ \\ 
$50$ & $0.200$ & $0.350$ & $27.050$ & $16.255$ & $\textbf{0.724}$ & $23.497$ & $\textbf{18.519}$ & $23.345$ \\ 
$100$ & $0.200$ & $0.350$ & $27.680$ & $16.782$ & $\textbf{0.341}$ & $22.484$ & $\textbf{18.023}$ & $23.375$ \\ 
$25$ & $0.500$ & $0.350$ & $139.265$ & $94.263$ & $\textbf{11.670}$ & $122.559$ & $91.154$ & $\textbf{48.796}$ \\ 
$50$ & $0.500$ & $0.350$ & $146.536$ & $97.070$ & $\textbf{3.707}$ & $122.424$ & $89.468$ & $\textbf{46.108}$ \\ 
$100$ & $0.500$ & $0.350$ & $149.750$ & $98.401$ & $\textbf{1.424}$ & $121.385$ & $87.669$ & $\textbf{44.767}$ \\ 
$25$ & $0.800$ & $0.350$ & $262.446$ & $233.520$ & $\textbf{54.512}$ & $257.264$ & $236.381$ & $\textbf{155.683}$ \\ 
$50$ & $0.800$ & $0.350$ & $265.817$ & $234.813$ & $\textbf{25.930}$ & $256.205$ & $234.103$ & $\textbf{143.190}$ \\ 
$100$ & $0.800$ & $0.350$ & $269.327$ & $237.345$ & $\textbf{15.345}$ & $255.937$ & $233.731$ & $\textbf{138.254}$ \\ 
\hline \\[-1.8ex] 
\end{tabular}  
  \end{table}

   \begin{table}[!htbp] \centering \footnotesize
  \caption{Average empirical Cram\'er--von Mises distances (non-normalized) between the CDFs of the estimated complete data GEV and the proposed modelling approach computed over $1000$ replications (``mle-obs'' - observed likelihood, ``mle-uncond'' - unconditional weighted likelihood, ``mle-cond'' - conditional weighted likelihood'', ``mom-obs'' - observed moments, ``mom-uncond'' - unconditional weighted moments, ``mom-cond'' - conditional weighted moments) for MCAR data generated from an t distribution (df=5) with blocks containing $100$ observations. Simulation parameters: ``sims'' - number of blocks, ``pbm'' - proportion of blocks with missing observations, ``pm'' - expected proportion of missing observations within blocks.} 
    \label{simtablet-MCAR} 
\begin{tabular}{@{\extracolsep{5pt}} ccc|ccc|ccc} 
\\[-1.8ex]\hline 
\hline \\[-1.8ex] 
sims & pbm & pm & mle-obs & mle-uncond & mle-cond & mom-obs & mom-uncond & mom-cond \\
\hline \\[-1.8ex] 
$25$ & $0.200$ & $0.050$ & $0.081$ & $0.079$ & $\textbf{0.073}$ & $\textbf{2.383}$ & $2.426$ & $2.623$ \\ 
$50$ & $0.200$ & $0.050$ & $0.051$ & $0.048$ & $\textbf{0.039}$ & $\textbf{1.843}$ & $1.871$ & $2.103$ \\ 
$100$ & $0.200$ & $0.050$ & $0.028$ & $0.027$ & $\textbf{0.020}$ & $\textbf{1.616}$ & $1.640$ & $1.891$ \\ 
$25$ & $0.500$ & $0.050$ & $0.269$ & $0.258$ & $\textbf{0.217}$ & $\textbf{2.420}$ & $2.504$ & $3.068$ \\ 
$50$ & $0.500$ & $0.050$ & $0.143$ & $0.137$ & $\textbf{0.094}$ & $\textbf{1.867}$ & $1.929$ & $2.636$ \\ 
$100$ & $0.500$ & $0.050$ & $0.094$ & $0.090$ & $\textbf{0.047}$ & $\textbf{1.562}$ & $1.605$ & $2.325$ \\ 
$25$ & $0.800$ & $0.050$ & $0.507$ & $0.500$ & $\textbf{0.373}$ & $\textbf{2.412}$ & $2.485$ & $3.466$ \\ 
$50$ & $0.800$ & $0.050$ & $0.306$ & $0.298$ & $\textbf{0.160}$ & $\textbf{1.918}$ & $1.964$ & $3.157$ \\ 
$100$ & $0.800$ & $0.050$ & $0.219$ & $0.213$ & $\textbf{0.080}$ & $\textbf{1.435}$ & $1.478$ & $2.703$ \\ 
$25$ & $0.200$ & $0.200$ & $0.569$ & $0.484$ & $\textbf{0.365}$ & $\textbf{2.460}$ & $3.148$ & $4.582$ \\ 
$50$ & $0.200$ & $0.200$ & $0.381$ & $0.297$ & $\textbf{0.168}$ & $\textbf{1.812}$ & $2.447$ & $4.069$ \\ 
$100$ & $0.200$ & $0.200$ & $0.245$ & $0.184$ & $\textbf{0.072}$ & $\textbf{1.549}$ & $2.130$ & $3.806$ \\ 
$25$ & $0.500$ & $0.200$ & $1.957$ & $1.672$ & $\textbf{0.879}$ & $\textbf{3.268}$ & $4.566$ & $8.314$ \\ 
$50$ & $0.500$ & $0.200$ & $1.505$ & $1.229$ & $\textbf{0.443}$ & $\textbf{2.415}$ & $3.614$ & $7.889$ \\ 
$100$ & $0.500$ & $0.200$ & $1.170$ & $0.934$ & $\textbf{0.194}$ & $\textbf{1.725}$ & $2.888$ & $7.264$ \\ 
$25$ & $0.800$ & $0.200$ & $4.188$ & $3.889$ & $\textbf{1.498}$ & $\textbf{4.575}$ & $5.608$ & $11.160$ \\ 
$50$ & $0.800$ & $0.200$ & $3.294$ & $2.969$ & $\textbf{0.664}$ & $\textbf{3.526}$ & $4.612$ & $10.996$ \\ 
$100$ & $0.800$ & $0.200$ & $2.856$ & $2.581$ & $\textbf{0.318}$ & $\textbf{2.737}$ & $3.692$ & $10.340$ \\ 
$25$ & $0.200$ & $0.350$ & $1.364$ & $0.923$ & $\textbf{0.611}$ & $\textbf{3.110}$ & $5.653$ & $8.083$ \\ 
$50$ & $0.200$ & $0.350$ & $0.940$ & $0.575$ & $\textbf{0.268}$ & $\textbf{2.108}$ & $4.686$ & $7.284$ \\ 
$100$ & $0.200$ & $0.350$ & $0.761$ & $0.422$ & $\textbf{0.136}$ & $\textbf{1.621}$ & $4.132$ & $6.864$ \\ 
$25$ & $0.500$ & $0.350$ & $5.461$ & $3.791$ & $\textbf{1.646}$ & $\textbf{5.560}$ & $10.582$ & $15.272$ \\ 
$50$ & $0.500$ & $0.350$ & $4.382$ & $2.877$ & $\textbf{0.773}$ & $\textbf{4.353}$ & $9.254$ & $14.884$ \\ 
$100$ & $0.500$ & $0.350$ & $3.900$ & $2.429$ & $\textbf{0.347}$ & $\textbf{3.673}$ & $8.685$ & $14.566$ \\ 
$25$ & $0.800$ & $0.350$ & $12.059$ & $10.000$ & $\textbf{2.803}$ & $\textbf{10.967}$ & $16.135$ & $21.133$ \\ 
$50$ & $0.800$ & $0.350$ & $10.506$ & $8.575$ & $\textbf{1.166}$ & $\textbf{9.213}$ & $14.336$ & $20.142$ \\ 
$100$ & $0.800$ & $0.350$ & $9.515$ & $7.675$ & $\textbf{0.602}$ & $\textbf{7.960}$ & $12.955$ & $19.439$ \\ 
\hline \\[-1.8ex] 
\end{tabular}  
  \end{table}

    \begin{table}[!htbp] \centering \footnotesize
  \caption{Average empirical Cram\'er--von Mises distances (non-normalized) between the CDFs of the estimated complete data GEV and the proposed modelling approach computed over $1000$ replications (``mle-obs'' - observed likelihood, ``mle-uncond'' - unconditional weighted likelihood, ``mle-cond'' - conditional weighted likelihood'', ``mom-obs'' - observed moments, ``mom-uncond'' - unconditional weighted moments, ``mom-cond'' - conditional weighted moments) for MAR data generated from an t distribution (df=5) with blocks containing $50$ or $100$ observations. Simulation parameters: ``sims'' - number of blocks, “apm” - average proportion of missingness across the series.} 
\label{simtablet-MAR}
\begin{tabular}{@{\extracolsep{5pt}} ccc|ccc|ccc} 
\\[-1.8ex]\hline 
\hline \\[-1.8ex] 
sims & apm & n & mle-obs & mle-uncond & mle-cond & mom-obs & mom-uncond & mom-cond \\ 
\hline \\[-1.8ex] 
$25$ & $0.050$ & $50$ & $0.784$ & $0.691$ & $\textbf{0.508}$ & $\textbf{2.624}$ & $3.218$ & $4.634$ \\ 
$50$ & $0.050$ & $50$ & $0.468$ & $0.396$ & $\textbf{0.218}$ & $\textbf{1.852}$ & $2.325$ & $3.864$ \\ 
$100$ & $0.050$ & $50$ & $0.354$ & $0.283$ & $\textbf{0.098}$ & $\textbf{1.366}$ & $1.807$ & $3.425$ \\ 
$25$ & $0.150$ & $50$ & $4.227$ & $2.689$ & $\textbf{1.380}$ & $\textbf{5.098}$ & $8.597$ & $12.216$ \\ 
$50$ & $0.150$ & $50$ & $3.528$ & $2.073$ & $\textbf{0.635}$ & $\textbf{3.768}$ & $7.223$ & $11.147$ \\ 
$100$ & $0.150$ & $50$ & $2.981$ & $1.636$ & $\textbf{0.280}$ & $\textbf{3.032}$ & $6.479$ & $10.694$ \\ 
$25$ & $0.250$ & $50$ & $11.616$ & $5.537$ & $\textbf{2.376}$ & $\textbf{11.547}$ & $15.625$ & $19.083$ \\ 
$50$ & $0.250$ & $50$ & $10.000$ & $4.292$ & $\textbf{1.022}$ & $\textbf{9.573}$ & $14.420$ & $18.581$ \\ 
$100$ & $0.250$ & $50$ & $9.138$ & $3.728$ & $\textbf{0.515}$ & $\textbf{8.540}$ & $13.787$ & $18.219$ \\ 
$25$ & $0.050$ & $100$ & $0.658$ & $0.570$ & $\textbf{0.424}$ & $\textbf{2.516}$ & $3.134$ & $4.843$ \\ 
$50$ & $0.050$ & $100$ & $0.524$ & $0.442$ & $\textbf{0.230}$ & $\textbf{1.827}$ & $2.405$ & $4.362$ \\ 
$100$ & $0.050$ & $100$ & $0.356$ & $0.288$ & $\textbf{0.104}$ & $\textbf{1.494}$ & $2.021$ & $4.138$ \\ 
$25$ & $0.150$ & $100$ & $4.507$ & $2.862$ & $\textbf{1.314}$ & $\textbf{5.225}$ & $9.196$ & $13.890$ \\ 
$50$ & $0.150$ & $100$ & $3.336$ & $1.969$ & $\textbf{0.603}$ & $\textbf{3.626}$ & $7.990$ & $13.194$ \\ 
$100$ & $0.150$ & $100$ & $2.980$ & $1.672$ & $\textbf{0.286}$ & $\textbf{3.022}$ & $7.435$ & $12.826$ \\ 
$25$ & $0.250$ & $100$ & $11.693$ & $5.864$ & $\textbf{2.411}$ & $\textbf{11.259}$ & $16.811$ & $21.642$ \\ 
$50$ & $0.250$ & $100$ & $9.854$ & $4.406$ & $\textbf{1.059}$ & $\textbf{9.228}$ & $15.859$ & $21.019$ \\ 
$100$ & $0.250$ & $100$ & $9.108$ & $3.843$ & $\textbf{0.471}$ & $\textbf{8.248}$ & $15.254$ & $20.724$ \\ 
\hline \\[-1.8ex] 
\end{tabular} 
  \end{table}

\section{Tidal Wave Surges in Eastern Canada}\label{sec:data}

\begin{figure}
    \subfloat{\includegraphics[width=\textwidth]{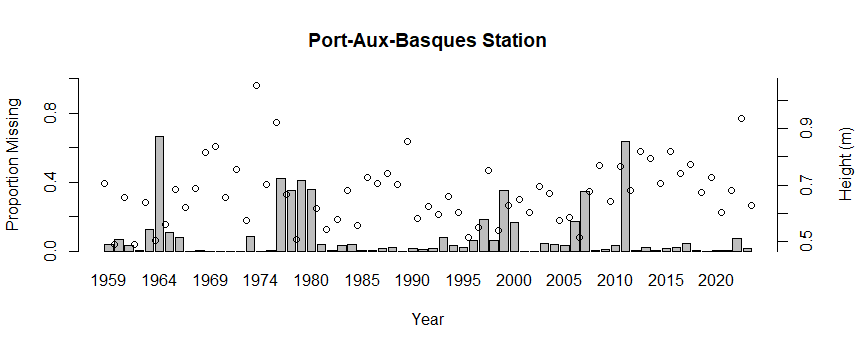}} 
    
    \subfloat{\includegraphics[width=\textwidth]{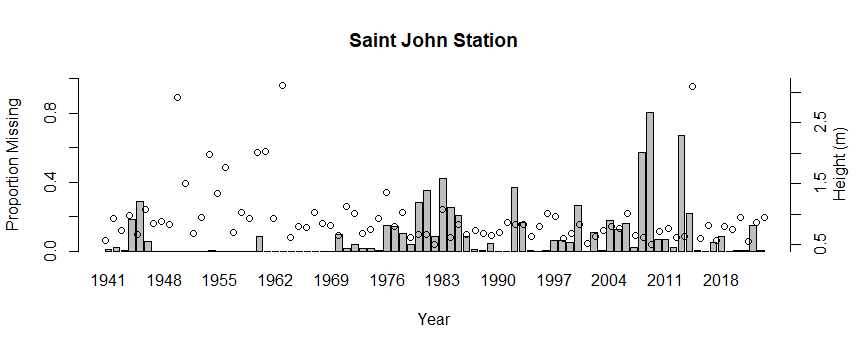}} 
    
    \subfloat{\includegraphics[width=\textwidth]{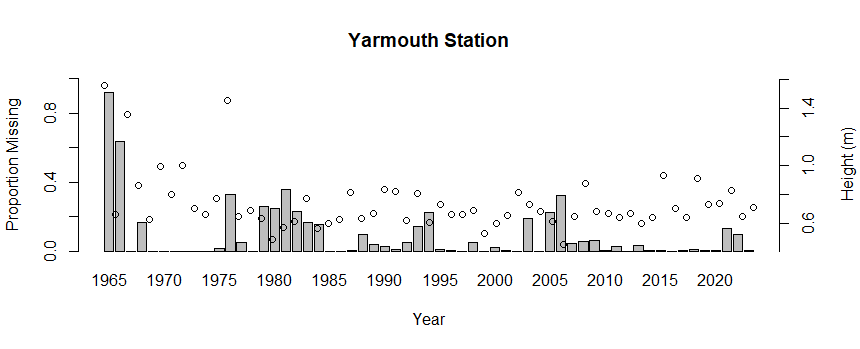}}
    \caption{Barplots of the proportion of missing observations measured by year and scatterplots of the observed maxima for each year for the Saint John (1941-2023), Yarmouth (1965-2023) and Port-aux-Basques (1959-2023) stations.}
    \label{DataAppPlot}
\end{figure}

In this data application, we modeled extremal wave surges to predict wave surge return levels. We used hourly water level data from three locations in Atlantic Canada: Saint John (ref \#: 65), Yarmouth (ref \#: 365) and Port-Aux-Basques (ref \#: 665) with corresponding time periods of 1941-2023 (83 years), 1965-2023 (59 years) and 1959-2023 (64 years), respectively. The number of years with missing observations for each location were 60, 45, and 56 with corresponding mean proportions of missing data in years with missing observations of $12.11\%$, $12.02\%$ and $10.67\%$, respectively. The analyzed data are publicly available and can be accessed at \url{https://www.tides.gc.ca/}. For a visual representation of the proportion of missing observations by year across the three stations and the corresponding observed maxima for each year, see Figure \ref{DataAppPlot}.

We preprocessed the data to remove the deterministic tidal effects using the TideHarmonics $\mathtt{R}$ package to obtain the wave surge levels by fitting the following model for the tidal effect:
$$T(t) = M(t) + \sum_{n=1}^{N} A_n \cos\left[ \pi\left\{\omega_n t - \psi_n\right\}/180 \right],$$
where $M(t)$ is the time-varying mean sea-level and the summation expression corresponds to the regular high/low tide pattern in the waves \cite{steph}. After determining an estimator of $T(t)$ at each recorded time $t$, we subtracted the estimated tide component from the observed wave heights to obtain an estimate of the stochastic wave surge series. We calculated the 20-, 50- and 100-year return levels for the wave surges using the proposed estimation procedures from Section \ref{sec:WE}. The approximate standard errors for each method were calculated using the nonparametric bootstrap procedure of \cite{efron} by resampling the observed maxima and weight pairs with replacement. 

The three stations selected for this data analysis illustrate different comparisons between estimation methods. We compared the proposed weighted estimation procedures to the estimation procedures which assumed all maxima are correctly observed, only the complete blocks are observed, and only the blocks with under 10\% missingness are observed. We note that for the case of the complete blocks only, we required the bootstrap samples to contain at least 4 observations each to ensure the GEV parameters and parameter estimator standard errors could be computed. The results are reported in Table \ref{locations}.

For Saint John, both conditional weighted approaches yield much larger return level estimates, with the moment-based approach having by far the largest return level estimates. However the standard error associated with these methods has approximately doubled, which is a reflection of the larger return level estimates and may also reflect the uncertainty associated with the observed maxima. The unconditional weighted approaches yield slightly larger return level estimates than for the observed estimation approaches with similar levels of uncertainty. For the Yarmouth and Port-Aux-Basques locations, the difference between results is much more moderate, with very little difference between the likelihood-based approaches at the Port-Aux-Basques location.

There were also interesting comparisons of the estimation methods with complete data and the 10\% missing complete data analyses. For these datasets, it is apparent from the large standard errors that using only complete blocks for ML estimation is not a good strategy as this will severely reduce the effective sample size. A moderate increase in the standard errors was found for the moments-based estimator using only complete cases, but not to the same degree as the ML method. This result can most likely be attributed to the PWM estimators' decent small sample behaviour \cite{beirlant2004}. When blocks missing more than 10\% of observations were removed from the analyses, both ML and PWM performed similarly to their observed data counterparts, i.e., analyses that ignore missing values, but had slightly larger standard errors due to the reduction in sample size.

\begin{table}[!htbp] \centering \footnotesize
\caption{Estimated 20-, 50-, and 100-year return levels and associated standard errors of surge heights (metres) measured at Saint John, Yarmouth and Port-Aux-Basques locations computed using various procedures (``MLE-obs'' - observed likelihood, ``MLE-uncond'' - unconditional weighted likelihood, ``MLE-cond'' - conditional weighted likelihood'', ``MoM-obs'' - observed moments, ``MoM-uncond'' - unconditional weighted moments, ``MoM-cond'' - conditional weighted moments, ``Complete" - only complete blocks used, ``Complete 10\%" - only blocks with under 10\% missingness used)}  
\begin{tabular}{@{\extracolsep{5pt}} c|ccc|ccc}
& \multicolumn{3}{c}{Return Levels Estimates} & \multicolumn{3}{c}{Standard Errors} \\ \hline 
Location & \multicolumn{6}{c}{Saint John} \\ \hline
Method & 20 years & 50 years & 100 years & 20 years & 50 years & 100 years \\ \hline
MLE-Obs & 1.7326 & 2.3842 & 3.0550 & 0.2411 & 0.4925 & 0.8182 \\
MLE-Uncond & 1.7599 & 2.4363 & 3.1384 & 0.2516 & 0.5193 & 0.8641 \\
MLE-Cond & 2.2255 & 3.3119 & 4.5133 & 0.4916 & 1.1007 & 1.9620 \\
MoM-Obs & 1.7470 & 2.4627 & 3.2261 & 0.2014 & 0.3729 & 0.5813 \\
MoM-Uncond & 1.8949 & 2.5679 & 3.2195 & 0.2230 & 0.4007 & 0.6074  \\ 
MoM-Cond & 3.3998 & 4.8085 & 6.0791 & 0.4276 & 0.8186 & 1.2522  \\ \hline
MLE-Complete & $2.443$ & $3.564$ & $4.761$ & $1.16 \times 10^{9}$ & $9.15 \times 10^{12}$ & $7.62 \times 10^{15}$ \\  
MLE-Complete 10\% & 1.733 & 2.384 & 3.055 & 0.2472 & 0.5103 & 0.8518 \\
MOM-Complete & 2.383 & 3.488 & 4.675 & 0.4724 & 0.7094 & 0.9812 \\ 
MOM-Complete 10\% & 1.747 & 2.463 & 3.226 & 0.2041 & 0.3760 & 0.5879 \\\hline \hline
Location & \multicolumn{6}{c}{Yarmouth} \\ \hline
Method & 20 years & 50 years & 100 years & 20 years & 50 years & 100 years \\ \hline
MLE-Obs & 1.0725 & 1.2374 & 1.3737 & 0.09246 & 0.1576 & 0.2313 \\
MLE-Uncond & 1.0163 & 1.1449 & 1.2471 & 0.07070 & 0.1147 & 0.1627 \\
MLE-Cond & 1.1306 & 1.3591 & 1.5703 & 0.1200 & 0.2158 & 0.3272  \\
MoM-Obs & 1.0970 & 1.3372 & 1.5644 & 0.09456 & 0.1673 & 0.2511 \\
MoM-Uncond & 1.0587 & 1.1340 & 1.1799 & 0.08093 & 0.1154 & 0.1445  \\ 
MoM-Cond & 1.4550 & 1.5918 & 1.6714 & 0.1688 & 0.2321 & 0.2782 \\ \hline
MLE-Complete & $1.164$ & $1.445$ & $1.714$ & $3.57 \times 10^{10}$ & $1.03 \times 10^{15}$ & $2.30 \times 10^{18}$ \\  
MLE-Complete 10\% & 1.073 & 1.237 & 1.374 & 0.0922 & 0.1588 & 0.2352 \\
MOM-Complete & 1.174 & 1.514 & 1.865 & 0.1772 & 0.2698 & 0.3848 \\ 
MOM-Complete 10\% & 1.097 & 1.3372 & 1.564 & 0.09421 & 0.1680 & 0.2530 \\ \hline \hline
Location & \multicolumn{6}{c}{Port-Aux-Basques} \\ \hline
Method & 20 years & 50 years & 100 years & 20 years & 50 years & 100 years \\ \hline
MLE-Obs & 0.8790 & 0.9500  & 1.0007 & 0.03292 & 0.04954 & 0.06567  \\
MLE-Uncond & 0.8799  & 0.9499 & 0.9997  & 0.03393 & 0.05099 & 0.06736  \\
MLE-Cond & 0.8965  & 0.9632 & 1.0100 & 0.03533 & 0.05217 & 0.06804 \\
MoM-Obs & 0.8790 & 0.9469 & 0.9945 & 0.03355 & 0.05011 & 0.06573 \\
MoM-Uncond & 0.9261 & 0.9690 & 0.9925 & 0.03910 & 0.05125 & 0.06101 \\ 
MoM-Cond & 1.1137 & 1.1563  & 1.1759 & 0.07200 & 0.08899 & 0.09913  \\ \hline
MLE-Complete & $1.045$ & $1.154$ & $1.236$ & $2.38 \times 10^{26}$ & $1.490 \times 10^{36}$ & $3.272 \times 10^{43}$ \\  
MLE-Complete 10\% & 0.8790 & 0.9500 & 1.001 & 0.03407 & 0.0515 & 0.0683 \\
MOM-Complete & 1.072 & 1.181 & 1.261 & 0.1075 & 0.161 & 0.230 \\ 
MOM-Complete 10\% & 0.8790 & 0.9470 & 0.9945 & 0.0344 & 0.0515 & 0.0675 \\\hline \hline

\end{tabular}
\label{locations}
\end{table}

\section{Discussion}\label{sec:disc}

It can be challenging to account for missing observations in an extreme value analysis due to the magnitude of extreme records as well as the various ways in which observations (extreme or non-extreme) may be missing. It is critical that the proposed estimation method accounts for both the extremes and missing observations to ensure unbiased parameter estimates. This notion was reflected in the simulation study where there was a marked improvement in performance between the weighted and unweighted likelihood estimators. Although the simulation of missing values may seem unrealistic to occur exactly in real data, they provide a spectrum of possible missingness scenarios from which we believe missingness in environmental applications would occur. For this reason, the proposed weighted estimation methods were also applied to water level data collected from tidal gauges on the Eastern coast of Canada. Our estimates indicated different degrees of variation in the return level estimates from the proposed methods across locations which possibly reflects differences in the missingness mechanisms between the locations. 
\

While the proposed weighted estimators are straightforward in their implementation and application to simulated or real-world data sets, there are still a variety of open problems related to missing observations in the context of extreme value analysis. First, in our proposed analyses, we conducted inferences on the unknown GEV parameters under a block maxima framework conditionally on the number of observations (missing and non-missing) within each block. While such an assumption can simplify the resulting analyses, we acknowledge that there could be a loss of efficiency in the resulting estimators. To conduct an unconditional analysis, assumptions are required on the independence and distribution of the series level data as well as the specific missingness mechanism to estimate the count of missing observations within any particular block. As the asymptotic GEV distribution does not directly depend (in terms of parameter estimation) on the specific form of the data generating distribution, it is unclear how inferences on the series level data distribution can be used to improve inferences on the GEV block maxima distribution. In future work, other weights could be considered which account for levels of missingness but reduce uncertainty in a univariate extreme value analysis. Another problem is the incorporation of missing observations in an extreme value analysis which includes a spatial component. For example, in our analysis of wave surges, the surge levels across stations are necessarily dependent in addition to a potential dependence between the stations' missing data mechanisms. Dupuis and Morgenthaler considered weighted maximum likelihood estimators for bivariate extreme value problems, but it is unclear which form of weights should be utilized in a spatial setting \cite{dupui}. Applications of extreme value analysis including spatial components can be found in the works of \cite{muis} and \cite{ross}.

\appendix

\section{Technical Remarks}

\textbf{Weighted Maximum Likelihood}

Using the relevance maximum likelihood setup of \cite{hu}, we denote the observed block maxima by $M_{n_1}, ..., M_{n_k}$ and assume that they are independently drawn from densities $f_j$, for $j = 1, ..., k$. As the observed maxima are not necessarily drawn directly from the GEV, given by $f(\cdot; \theta)$ for $\theta \in \Omega$, we introduce relevance weights to weight the relevance of density $f_j$ to the target density $f(\cdot; \theta)$. The weights are normalized as $\frac{\hat{w}_i}{\sum_{j=1}^{k} \hat{w}_j}$ for $i = 1, ..., k$ to ensure they are non-negative and they sum to $1$. Changing from the original weights $\hat{w}_i$ to the normalized weights results in the same maximum likelihood estimators. 

One of the main challenges in establishing the asymptotic properties of the weighted ML estimator is defining the form of the observed maxima densities, $f_j$, for $j = 1, ..., k$ in conjunction with their corresponding weighting functions. Specifically, in a standard extreme value analysis without missing observations where the block size is assumed to be large, the density of the observed maximum can be approximated by the GEV distribution. In the presence of missing observations, the observed maximum may not correspond to the true block maximum but could correspond to a lower order statistic. The way in which the data are assumed to be missing will directly impact the asymptotic distribution of the observed block maxima. For example, when the upper 10\% of the series level data are always missing according to
an MNAR missingness mechanism, the limiting behaviour of the observed block maximum is not immediately clear as its order will change as block size goes to infinity. The rate the order changes which depends on the missingness mechanism. Thus, the observed maxima densities must be uniquely defined for each type of missingness mechanism. 

A second difficulty in establishing the asymptotic properties for the weighted maximum likelihood estimator is through the proposed weighting functions. The proposed weights are both random in that they either depend on the random proportion of observed data within a block (i.e. the unconditional weights) or directly depend on the series level data through the empirical cumulative distribution function. We note however that the relevance weighted maximum likelihood estimator described in \cite{hu} assumes the relevance weights $p_{nj}$ are not random and are fixed by the researcher. As the asymptotic proof approach is based on the original techniques of Wald and Cramer, it is not clear how the approach can be adjusted to allow for random relevance weighting functions.  

\textbf{Weighted Probability-Weighted Moments Estimator}

To establish asymptotic Normality of the Probability-Weighted Moment Estimator, \cite{hoski} appealed to the proof technique of \cite{chern} which required three conditions. We will verify that the three conditions still hold with the inclusion of the unconditional weights (assuming the number of missing/observed series level data per block are fixed). For specific details on the conditions and how they relate to more general estimation problems, refer to the original work of \cite{chern}. First, condition A states that the terms inside the summation are continuously differentiable. As the inclusion of constant weighting functions does not affect differentiability or the unweighted estimator proposed in \cite{hoski}, this condition holds trivially. Similarly, conditions B and C state that particular summations are expressions that grow at little o rates. However, since all the unconditional weights are constant and bounded above by $1$, then these conditions will still hold as the same upper bound may be assumed. In the case where the weights are conditionally based on the observed maxima, the original conditions of \cite{chern} may not be sufficient as the differentiability condition will depend on the conditional weighting function and the rates of growth for the summations will also depend on the inputs of the conditional weighting functions. 

\bibliographystyle{chicago}
\bibliography{wileyNJD-AMA}

\section*{Author contributions}
All listed authors contributed equally to this manuscript.

\section*{Acknowledgments}
This work was supported by the Natural Sciences Engineering Research Council of Canada Discovery Grant Program (JHM and OAM).

\section*{Financial disclosure}

The authors declare no financial disclosures.

\section*{Conflict of interest}

The authors declare no potential conflict of interests.

\section*{Supporting information}

\comR{Supplementary materials available upon request.}




\end{document}